\documentclass[11pt,a4paper]{article}
\usepackage{times}
\usepackage{tgtermes}
\usepackage[T1]{fontenc}
\usepackage[utf8]{inputenc}
\usepackage[french]{babel}
\usepackage{fancyhdr}
\usepackage{setspace}

\usepackage{amssymb}
\usepackage{undertilde}
\usepackage{amsmath}
\usepackage{graphicx}
\usepackage{wrapfig}

\usepackage{titlesec}
\titleformat{\section}
  {\normalfont\fontsize{16}{20}\bfseries}{\thesection}{1em}{}
\titleformat{\subsection}
  {\normalfont\fontsize{16}{20}\bfseries}{\thesubsection}{1em}{}
\titlespacing*{\section}
{0pt}{2.ex plus 1ex minus .2ex}{.0ex}
\titlespacing*{\subsection}
{0pt}{0.ex}{.0ex}

\begin{document}

\begin{center}
\begin{spacing}{2.05}
{\fontsize{20}{20}
\bf
Validation d'un modèle pour un polymère électroactif}
\end{spacing}
\end{center}
\vspace{-1.25cm}
\begin{center}
{\fontsize{14}{20}
\bf
M. TIXIER\textsuperscript{a}, J. POUGET\textsuperscript{b}\\
\bigskip
}
{\fontsize{12}{20}
a. Département de Physique, Université de Versailles Saint Quentin,
45, avenue des Etats-Unis, F-78035 Versailles, France;
mireille.tixier@uvsq.fr\\
b. Sorbonne Université, CNRS, Institut Jean le Rond d'Alembert, UMR 7190, F-75005 Paris, France; pouget@lmm.jussieu.fr\\
}
\end{center}

\vspace{10pt}

{\fontsize{16}{20}
\bf
R\'esum\'e :
}
\medskip

\textit{Les polymères électro-actifs (PEA) ioniques sont susceptibles de se déformer sous l'action d'un champ électrique, ce qui leur confère de nombreuses applications comme capteur, actionneur ou récupérateur d'énergie.\newline}
\textit{Un composite métal-polymère ionique (IPMC) est constitué d'un film de polymère ionique recouvert sur ses deux faces d'une fine couche de métal servant d'électrodes. Le polymère est saturé d'eau, ce qui entraîne sa dissociation quasi complète ; les ions négatifs restent attachés aux chaînes polymères qui peuvent être considérées comme un milieu poreux dans les pores duquel se déplacent l'eau et les cations. Sous l'effet d'un champ électrique orthogonal à la lame, les cations migrent vers l'électrode négative, entraînant avec eux une partie de l'eau par un phénomène d'osmose et provoquant la flexion de la lame.\newline}
\textit{Nous avions préalablement établi les lois de conservation et de comportement de ce matériau. Ce modèle a été appliqué au cas d'une lame de PEA encastrée - libre soumise à une différence de potentiel continue entre ses deux faces (cas statique). L'amplitude de la flexion étant grande, les efforts appliqués et la flèche sont calculés en utilisant un modèle de poutre en grands déplacements. Nous avons également étudié le cas où une force de blocage empêche l'extrémité libre de se déplacer. La permittivité du milieu étant susceptible d'augmenter avec la concentration en cations, nous avons comparé plusieurs modèles de permittivité : fonctions constante, linéaire ou affine.\newline}
\textit{Les simulations numériques ont été effectuées dans le cas du Nafion. La résolution du système d'équations nous a permis de tracer les profils des différentes grandeurs (concentration en cations, potentiel et induction électriques, pression), qui s'avèrent très raides au voisinage des électrodes.
Les valeurs obtenues pour la flèche et de la force de blocage sont en bon accord avec les données expérimentales publiées dans la littérature. Nous avons aussi étudié l'influence de la géométrie de la lame, identique pour les trois modèles. La variation de ces deux grandeurs avec le potentiel électrique imposé dépend en revanche du modèle de permittivité choisi, ce qui permet de les discriminer.}

\vspace{20pt}

{\fontsize{16}{20}
\bf
Abstract :
}
\bigskip

\textit{Ionic electro-active polymers (EAP) are able to deform under the action of an electric field, which confers them many applications as sensor, actuator or energy recovery.\newline}
\textit{An ionic metal-polymer composite (IPMC) consists in an ionic polymer film coated on both sides with a thin layer of metallic electrodes. The polymer is saturated with water, which results in its quasi complete dissociation : negative ions remain bound to the polymer backbone, which can be considered as a porous medium, and small cations are released in water. When an electric field orthogonal to the strip is applied, the cations move towards the negative electrode and carry solvent away by osmosis, causing the bending of the strip.\newline}
\textit{We had previously established the conservation laws and the constitutive equations of this material. This model has been applied to the case of a cantilevered EAP strip subjected to a continuous voltage between its two faces (static case). Since the amplitude of the bending is large, the applied forces and the deflection are calculated using a beam model in large displacements. We also studied the case of a blocking force preventing the free end from moving. The material permittivity may increase with cations concentration, so we have compared several permittivity models : constant, linear and affine functions.\newline}
\textit{Numerical simulations were performed in the case of Nafion. The resolution of the equations system enabled us to draw the profiles of various quantities (cations concentration, electric potential and induction, pressure), which drastically vary near the electrodes.
The tip displacement and blocking force values obtained fit well the experimental data published in the literature. We also studied the influence of the strip geometry, which is identical for the three models. On the contrary, the variations of these two quantities with the imposed electric potential depend on the chosen permittivity model, which allows to discriminate them.
}

\vspace{28pt}

{\fontsize{14}{20}
\bf
Mots clefs : Electro-active polymers - Multiphysics coupling - Polymer mechanics - Nafion - Smart materials
}

\section{Introduction}
\medskip
Les polymères électroactifs (PEA) peuvent être utilisés comme actionneurs ou capteurs, ce qui permet des applications très prometteuses dans nombre de domaines : confection de micro-pompes, micromanipulateurs voire micro-robots, muscles artificiels, conception d'ailes battantes pour les micro-drones, récupération d'énergie....

Nous nous sommes plus particulièrement intéressés aux polymères ioniques de type Nafion ou Flemion sous forme de lames minces. Pour être actionné, le polymère doit être recouvert sur ses deux faces d'une fine couche de métal (or ou platine) servant d'électrodes; l'ensemble forme un IPMC. Il doit en outre être saturé en eau, ce qui provoque sa dissociation complète et la libération dans l'eau de cations de petite taille ($H^{+}$, $Li^{+}$ ou $Na^{+}$). Les anions restent fixés sur les chaînes polymères. Lorsque l'on applique un champ électrique orthogonal à la lame, les cations migrent vers l'électrode négative (cathode), entraînant avec eux l'eau par un phénomène d'osmose. Cette migration provoque un gonflement du polymère du côté de la cathode et une contraction sur l'autre face, entraînant une flexion de la lame vers l'anode. Ce phénomène met donc en jeu des couplages électro-mécano-chimiques. Le déplacement de l'extrémité libre est de l'ordre de quelques millimètres pour une lame de $200\;\mu m$ d'épaisseur et de quelques centimètres de long soumise à une différence de potentiel de quelques volts, avec un temps caractéristique d'une seconde environ \cite{nemat2000}.

Nous avions précédemment modélisé ce système grâce à la thermomécanique des milieux continus : les chaînes polymères chargées négativement sont assimilées à un milieu poreux déformable dans lequel s'écoule une solution ionique formée par l'eau et les cations. Les lois de conservation et de comportement obtenues sont rappelées dans le second paragraphe.
Nous avons appliqué ce modèle à la flexion d'une lame encastrée à l'une de ses extrémités, l'autre extrémité étant soit libre, soit maintenue fixe par un effort tranchant (force de blocage). La flèche étant importante dans le premier cas, il est nécessaire d'utiliser un modèle de poutre en grands déplacements qui est présenté au paragraphe 3.

Le système d'équations est ensuite simplifié, adimensionné et résolu aux paragraphes 4 et 5. La permittivité d'un matériau étant susceptible d'augmenter avec sa conductivité, nous avons envisagé trois modèles de permittivité : des fonctions constante, linéaire et affine de la concentration en cations.

Dans le paragraphe 6, nous exposons les résultats de nos simulations. Nous avons tracé les profils des différentes grandeurs  dans l'épaisseur de la lame, qui ont tous une allure similaire : constants dans la partie centrale, avec une forte variation au voisinage des bornes. Nous avons également étudié l'influence des caractéristiques géométriques de la lame : longueur, largeur et épaisseur, et du potentiel électrique imposé. Nous avons comparé les trois modèles de permittivité et confrontés les résultats aux données expérimentales publiées dans la littérature. Nos conclusions sont détaillées dans le paragraphe 7.

\section{Modélisation du polymère}

\subsection{Hypothèses}
\medskip
Pour modéliser le polymère électro-actif, nous avons utilisé une approche de type "milieu continu". Les chaînes polymères chargées négativement sont assimilées à un milieu poreux déformable, homogène et isotrope dans lequel se déplacent l'eau et les cations. Le système est donc composé de trois constituants mobiles les uns par rapport aux autres : le solide poreux, le solvant (l'eau) et les cations, et de deux phases. Les phases solide et liquide (eau + cations) sont séparées par une interface. On néglige la gravité et l'induction magnétique. Les différentes phases sont supposées incompressibles et la solution diluée. On admet en outre que les
déformations du solide sont petites. 

Nous nous sommes basés sur un modèle à gros grains développé pour les mélanges à deux constituants \cite{Ishii06}. On définit deux échelles : à l'échelle microscopique ($100 \; A^{\circ}$ environ), le volume élémentaire ne contient qu'une seule phase; à l'échelle macroscopique (de l'ordre du micron), on définit un volume élémentaire représentatif contenant les deux phases. Les équations de conservation sont tout d'abord écrites à l'échelle microscopique pour chaque phase et pour les interfaces. On en déduit les équations macroscopiques du matériau par un processus de moyenne utilisant des fonctions de présence pour chaque phase \cite{Tixier1}.

\subsection{Equations de bilan et lois de comportement}
\medskip
On obtient ainsi les équations de bilan de la masse, de la quantité de mouvement (2), de l'énergie, de la charge électrique et les équations de Maxwell (1) relatives au matériau complet \cite{Tixier1}. En particulier :
\begin{equation}
\begin{tabular}{lll}
$\overrightarrow{rot}\overrightarrow{E}=\overrightarrow{0}\qquad \qquad $ & $%
div\overrightarrow{D}=\rho Z\qquad \qquad $ & $\overrightarrow{D}%
=\varepsilon \overrightarrow{E}$%
\end{tabular} \label{maxwell}
\end{equation}
\begin{equation}
\rho \frac{D\overrightarrow{V}}{Dt}=\overrightarrow{div}\utilde{\sigma }%
+\rho Z\overrightarrow{E}  \label{CQ}
\end{equation}%
où $\rho$ est sa masse volumique, $Z$ sa charge électrique massique, $\varepsilon$ sa permittivité diélectrique, $\overrightarrow{E}$ le champ électrique, $\overrightarrow{D}$ l'induction électrique, $\overrightarrow{V}$ la vitesse et $\utilde{\sigma}$ le tenseur des contraintes.

La relation de Gibbs du matériau est obtenue de façon similaire. En combinant cette dernière relation avec les lois de conservation précédentes, on peut déterminer la fonction de dissipation du système \cite{Tixier2}. La thermodynamique des processus irréversibles linéaires permet alors d'identifier 
les flux et les forces généralisées associées et d'en déduire les lois de comportement du matériau. On obtient une loi rhéologique de type Kelvin-Voigt (\ref{rheo}), une loi de Fourier généralisée, une loi de Darcy généralisée (\ref{Darcy}) et une loi de Nernst-Planck (loi de Fick généralisée) (\ref{Nernst}) :
\begin{equation}
\utilde{\sigma }=\lambda \left( tr\utilde{\epsilon }\right) \utilde{1}+2G%
\utilde{\epsilon }+\lambda_{v}\left( tr \dot{\utilde{\epsilon }} \right)
 \utilde{1}+2\mu_{v}\dot{\utilde{\epsilon }} \label{rheo}
\end{equation}
\begin{equation}
\overrightarrow{V_{4}}-\overrightarrow{V_{3}}\simeq -\frac{K}{\eta\phi
_{4}}\left[ \overrightarrow{grad}p-\left(C F - \rho _{2}^{0} Z_{3}\right) 
\overrightarrow{E}\right] \label{Darcy}
\end{equation}%
\begin{equation}
\overrightarrow{V_{1}}=-\frac{D}{C}\left[ \overrightarrow{grad}C-\frac{%
Z_{1}M_{1}C}{RT}\overrightarrow{E}+\frac{Cv_{1}}{RT}\left( 1-\frac{M_{1}}{%
M_{2}}\frac{v_{2}}{v_{1}}\right) \overrightarrow{grad}p\right] +%
\overrightarrow{V_{2}} \label{Nernst}
\end{equation}%
où où $\utilde{\epsilon }$ désigne le tenseur des déformations, $\utilde{1}$ le tenseur identité, $\lambda$ le premier coefficient de Lamé, $G$ le module 
de cisaillement, $\lambda_{v}$ et $\mu_{v}$ des coefficients viscoélastiques, $\eta$ désigne la viscosité dynamique du solvant, $\phi_{4}$ la fraction volumique de la solution, $K$ la 
perméabilité intrinsèque du solide, $p$ la pression, $T$ la température absolue, $D$ le coefficient de diffusion de masse des cations, $C$ la concentration molaire en cations, $F=96487~C~mol^{-1}$ la constante de Faraday, $M_{k}$ la masse molaire, $v_{k}$ le volume molaire partiel et $R=8,31 J~K^{-1}$ la constante universelle des gaz parfaits. Les indices $1$, $2$, $3$ et $4$ sont relatifs respectivement aux cations, au solvant, au solide et à la solution (solvant + cations).

\section{Modèle de poutre en grands déplacements}
\medskip

\begin{wrapfigure}{l}{0.5\textwidth}
\includegraphics [width=\linewidth]{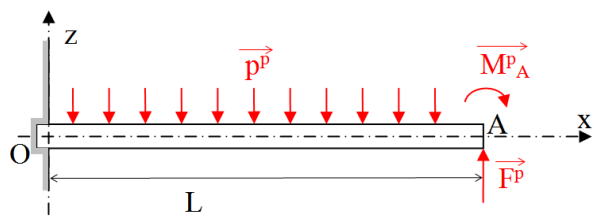}
\caption{Efforts exercés sur la poutre}
\label{fig:2}
\end{wrapfigure}

Considérons une lame de PEA de longueur $L$, de largeur $2l$ et d'épaisseur $2e$. Si elle est suffisamment mince, les efforts, contraintes et déplacements peuvent être déterminés en assimilant la lame à une poutre encastrée à son extrémité $O$. L'autre extrémité $A$ est soit libre, soit soumise à un effort tranchant $\overrightarrow{F^{p}}$. Lorsque l'on applique une différence de potentiel $\varphi_{0}$ entre les deux faces, les cations et le solvant se déplacent vers l'électrode négative, entraînant une variation de volume et un fléchissement de la lame.

Le système de forces appliqué à la poutre peut être modélisé par un effort réparti $\overrightarrow{p^{p}}$ et un moment fléchissant $\overrightarrow{M_{A}^{p}}$ appliqué à l'extrémité $A$.

Compte tenu des hypothèses précédentes, l'effort réparti est 
indépendant de la coordonnée $x$ et est orthogonal à la lame. Les forces électrostatiques intérieures à la lame ont une résultante nulle en raison de la condition d'électroneutralité. Pour la même raison, la force électrique générée par les électrodes s'annule également. On en déduit que l'effort réparti $\overrightarrow{p^{p}}$ est nul en tout $x$. 

Le moment fléchissant est exercé suivant l'axe $Oy$ et résulte des efforts de pression $p = \frac {\sigma_{xx}} {3}$ : 
\begin{equation}
M_{A}^{p}= \int_{-l}^{l} \int_{-e}^{e}\sigma_{xx}~z~dz~dy = 6l \int_{-e}^{e} p~z~dz
\end{equation}

\begin{wrapfigure}{r}{0.43\textwidth}
\includegraphics [width=\linewidth]{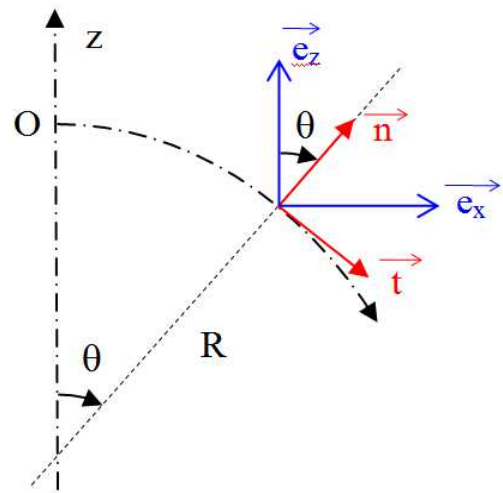}
\caption{Poutre en grands déplacements}
\label{fig:3}
\end{wrapfigure}

La flèche pouvant être très importante, il est nécessaire de faire un calcul en grands déplacements au moins dans le cas où l'extrémité $A$ est libre. Nous faisons les hypothèses habituelles pour le calcul des poutres : on admet que les sections droites de la lame restent planes et normales aux fibres après déformation (hypothèse de Bernoulli) et que les répartitions des contraintes et des déformations sont indépendantes des points d'application des forces extérieures (hypothèse de Barré Saint Venant).
On définit un repère local avec $\overrightarrow{t}$ et $\overrightarrow{n}$, vecteurs tangent et normal à la poutre; $s$ et $\overline{s}$ désignent les abscisses curvilignes respectivement le long de la poutre au repos et de la poutre déformée, $\overline{n}$ la coordonnée suivant la direction normale et $\theta$ l'angle de rotation d'une section droite. On choisit le point $O$ comme origine des abcisses curvilignes. Aucun effort normal n'étant appliqué, on admettra que la poutre ne s'allonge pas, donc que $d \overline{s}=ds$. Le moment de flexion $M^{p}$ en une section quelconque et le rayon de courbure $R^{p}$ s'écrivent :
\begin{equation}
M^{p} = F^{p} \left( L-x \right) + M_{A}^{p}
\qquad \qquad \qquad \frac{1}{R^{p}} = \frac{d \theta}{d \overline{s}}
\end{equation}

Choisissons un repère $Oxyz$ tel que l'axe $Oz$ soit orthogonal à la lame non déformée et l'axe $Ox$ suivant la longueur de la lame (cf figure 1). Soit $\overrightarrow{u}$, le champ de déplacements; son gradient vaut :
\begin{equation}
\utilde{Grad} \overrightarrow{u} = 
\begin{pmatrix}
   \left( 1 - \frac{\overline{n}}{R^{p}} \right) cos \theta - 1 & -sin \theta \\
   \left( 1 - \frac{\overline{n}}{R^{p}} \right) sin \theta & cos \theta - 1 
\end{pmatrix}
\end{equation}
On en déduit le tenseur des déformations :
\begin{equation}
\utilde{\epsilon} = \frac{1}{2} \left[ \left( \utilde{Grad} 
\overrightarrow{u} + \utilde{1} \right) ^{T} \left( 
\utilde{Grad} \overrightarrow{u} + \utilde{1} \right) 
- \utilde{1} \right]
\end{equation}
La poutre étant mince, $\left| \overline{n} \right| << R^{p}$. Il vient :
\begin{equation}
\epsilon_{xx} = - \frac{\overline{n}}{R^{p}} \left( 1 - \frac{\overline{n}}{2R^{p}} \right) \simeq - 
\frac{\overline{n}}{R^{p}}
\end{equation}
Dans le cas d'une poutre en flexion pure, la déformation vaut $\epsilon_{xx}=\frac{M^{p}}{EI^{p}} \overline{n}$ 
où $E$ est le module d'Young et $I^{p}=\frac{4le^{3}}{3}$ le moment quadratique par rapport à l'axe $Oy$. L'effort 
tranchant ayant un effet négligeable sur la flèche, on en déduit :
\begin{equation}
\begin{tabular}{ccc}
$\frac{1}{R^{p}} = \frac{d \theta}{d \overline{s}} = \frac{F^{p}}{E I^{p}}  \left( L-\overline{s} 
\right) - \frac{M_{A}^{p}}{E I^{p}} \qquad \qquad $ & soit
$\qquad \qquad \theta = \frac{F^{p}}{2E I^{p}} \overline{s} \left( 2L-\overline{s} \right) - 
\frac{M_{A}^{p}}{E I^{p}} \overline{s} $
\end{tabular}
\end{equation}
La flèche $w$ est alors obtenue en intégrant la relation $\frac{dz}{d \overline{s}} = sin \theta$.

Dans le cas d'une poutre encastrée libre ($F^{p}=0$), le rayon de courbure est constant; la poutre prend la forme 
d'un arc de cercle et la flèche à l'extrémité vérifie :
\begin{equation}
\begin{tabular}{cc}

$w = \frac{E I^{p}}{M_{A}^{p}} \left[ cos \left( \frac{M_{A}^{p}}{E I^{p}} L \right) -1 \right] \qquad \qquad $ & $\theta = - \frac{M_{A}^{p}}{E I^{p}} L $
\end{tabular}
\label{w}
\end{equation}
Le calcul en petits déplacements fournit :
\begin{equation}
w_{p} = -\frac{M_{A}^{p}}{2 E I^{p}} L^{2}
\end{equation}

Lorsqu'une force de blocage est exercée, la flèche s'annule à l'extrémité de la poutre. En petits déplacements, on obtient :
\begin{equation}
F^{p} = \frac{3 M_{A}^{p}}{2L}
\label{FP}
\end{equation}
En grands déplacements, $F^{p}$ vérifie :
\begin{equation}
w = \int_{0}^{L} sin \left[\frac{F^{p}}{2E I^{p}} x \left( 2L-x \right) - \frac{M_{A}^{p}}{E I^{p}} 
x \right]~dx = 0
\end{equation}
On peut évaluer cette intégrale en utilisant les fonctions de Fresnel; compte tenu des valeurs de $M_{A}^{p}$ fournies par les simulations, on peut montrer que l'erreur commise en utilisant la formule \ref{FP} est inférieure à $1 \%$.

Evaluons les variations de la fraction volumique $\phi_{4}$. Soit un petit élément de volume $dV$ situé à une distance $z$ de l'axe de la poutre. D'après l'hypothèse de Bernoulli, lorsque la poutre fléchit avec un rayon de courbure $R^{p}$, ce volume devient $\frac{\left| R^{p} \right|+z}{\left| R^{p} \right|} dV$. Le volume de la phase solide est invariable, et la variation de fraction volumique de la phase liquide est de l'ordre de $d\phi_{4}=\frac {\phi_{3} dz} {\left| R^{p} \right|}=(1-\phi_{4}) \frac{M^{p}}{EI^{p}}$. Numériquement, cette variation est inférieure à $0,3\%$ sur l'épaisseur de la poutre. On fera donc dans toute la suite l'hypothèse que la fraction volumique $\phi_{4}$ est constante.

\section{Système d'équations pour une lame en flexion}
\medskip
Nous avons appliqué le modèle résumé en section 2 au cas d'une lame de polymère électro-actif se déformant sous l'action d'une différence de potentiel $\varphi_{0}$ constante entre les deux électrodes (cas statique). Les vitesses des différents constituants et les dérivées partielles par rapport au temps sont donc nulles.

\begin{wrapfigure}{l}{0.6\textwidth}
\includegraphics [width=\linewidth]{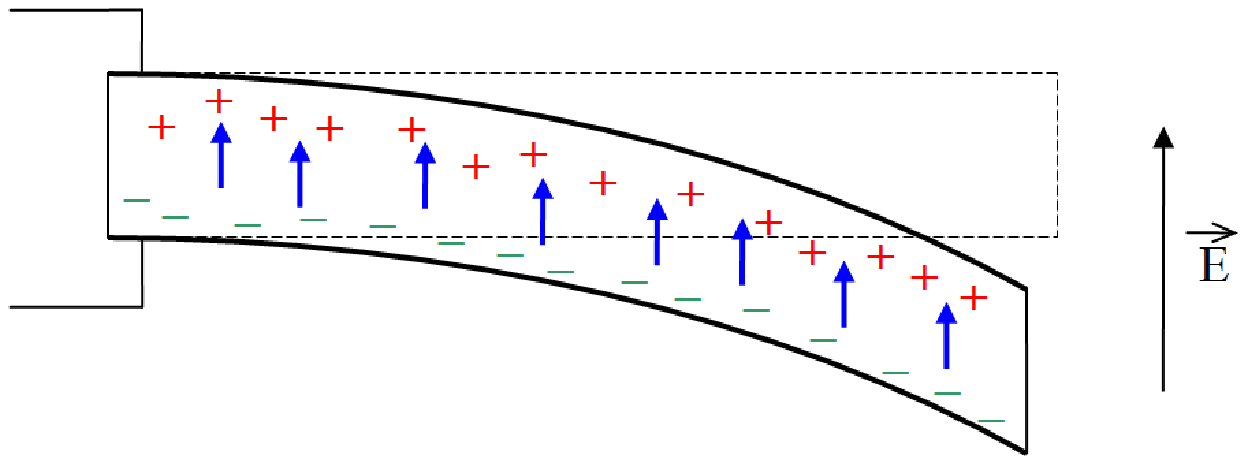}
\caption{Lame de PEA en flexion}
\label{fig:1}
\end{wrapfigure}

Nous avons choisi une lame de Nafion $Li^{+}$ de dimensions nominales $L=2~cm$, $e=100~\mu m$ et $l=2,5~mm$ soumise à une différence de potentiel $\varphi_{0} = 1~V$. La charge électrique massique des chaînes polymères est $Z_{3}\approx -9~10^{4}~C~kg^{-1}$ \cite{Collette} et la masse volumique de la phase solide $ \rho_{3}^{0} \approx 2078~kg~m^{-3}$ \cite{nemat2000}.
La fraction volumique de la solution $\phi_{4}$ est de l'ordre de $38\%$ \cite{nemat2000} et la masse volumique de l'eau ${\rho_{2}^{0}=1000~kg~m^{-3}}$, le module d'Young est $E=1,3 \; 10^{8} \; Pa$ \cite{Satterfield2009}. On choisit une température absolue ${T=300~K}$.

La permittivité diélectrique du matériau dépend fortement de sa conductivité, donc du nombre de charges électriques qu'il contient; elle dépend donc de la concentration en cations. Nous supposons qu'elle vérifie une loi de mélange :
\begin{equation}
\varepsilon = \varepsilon^{0} + \alpha C
\qquad \qquad avec \qquad \qquad
\alpha = \frac{\varepsilon_{moy}-\varepsilon^{0}}{C_{moy}}
\end{equation}

où $\varepsilon_{moy}$ désigne la permittivité moyenne du matériau et $C_{moy}$ sa concentration moyenne en cations. On peut considérer que les anions, attachés aux chaînes polymères, sont répartis de façon uniforme dans le matériau; l'électroneutralité permet d'écrire :
\begin{equation}
C_{moy}=-\frac{\left( 1-\phi _{4}\right) \rho _{3}^{0}Z_{3}}{\phi _{4}F}= 3082~mol~m^{-3}
\end{equation}

Nous avons envisagé trois modèles de permittivité : constante, linéaire et affine. Les valeurs numériques ont été ajustées de façon à obtenir une flèche et une force de blocage en accord avec les données de la littérature, soit $0,5<w<1,5 \; mm$ \cite{Nemat2002,NewburyTh} et $0,6<F^{p}<1,3 \; mN$ \cite{NewburyTh,Newbury2002,Newbury2003} :
\begin{equation}
\begin{tabular}{lll}
cas constant & $\varepsilon^{0} = 5 \;10^{-6}~F~m^{-1}$ &$\alpha=0$\\
cas linéaire & $\varepsilon^{0} = 0$ &$\alpha=10^{-3}~F~m^{-1}$\\
cas affine & $\varepsilon^{0} = 5 \;10^{-6}~F~m^{-1}$ &$\alpha=10^{-3}~F~m^{-1}$\\
\end{tabular}
\end{equation}
La permittivité moyenne a été mesurée par Deng et al \cite{Deng} pour un matériau très voisin du Nafion : $\varepsilon \sim 10^{-6}~F~m^{-1}$. D'autres auteurs \cite{Nemat2002,Farinholt} déduisent la permittivité de mesures de capacité et obtiennent une valeur de l'ordre de $10^{-3}~F~m^{-1}$, ce qui est l'ordre de grandeur des valeurs que nous obtenons dans les cas linéaire et affine.

Compte tenu des dimensions de la lame, le problème peut être considéré comme bidimensionnel dans le plan $Oxy$. En première approximation, on peut considérer que le champ et l'induction électriques sont parallèles à l'axe $Oz$ : $\overrightarrow{E}\simeq E_{z} \overrightarrow{e_{z}}$ et $\overrightarrow{D}\simeq D_{z} \overrightarrow{e_{z}}$. On admet en outre que $C$, $E_{z}$, $D_{z}$, $p$, le potentiel électrique $\varphi$ et la charge électrique volumique $\rho Z$ ne dépendent que de la variable $z$. On suppose enfin que le terme de pression de l'équation (\ref{Nernst}) est négligeable, hypothèse que nous vérifierons par la suite. Le système d'équations s'écrit alors :
\begin{equation}
\begin{tabular}{lll}
$E_{z}= - \frac {d \varphi} {dz} \qquad \qquad $ & 
$\frac {d D_{z}} {dz} = \rho Z \qquad \qquad $ & $D_{z} = \varepsilon E_{z}$ \\
$\frac {dp} {dz} = \left(C F - \rho _{2}^{0} Z_{3}\right) E_{z} \qquad \qquad $ &
$\frac {dC} {dz} = \frac{FC}{RT} E_{z} $ &$ $\\
\end{tabular} \label{Eqp}
\end{equation}%
où $\varepsilon$ est une fonction croissante de $C$ et :
\begin{equation}
\rho Z = \phi_{4} F \left(C - C_{moy} \right)
\end{equation}
Les conditions aux limites et la condition d'électroneutralité s'écrivent :
\begin{equation}
\begin{tabular}{ccc}
$\varphi(-e) =\varphi_{0}\qquad \qquad $ & $%
\varphi (e) = 0 \qquad \qquad $ & $
\int_{-e}^{e} \rho Z \, \mathrm{d}z =0 $
\end{tabular} \label{CL}
\end{equation}
Cette dernière condition équivaut à $D_{z}\left( e \right) = D_{z}\left( -e \right)$ d'après (\ref{Eqp}). 

Introduisons des variables adimensionnelles :
\begin{equation}
\begin{tabular}{cccc}
$\overline{E} = \frac {E_{z} e} {\varphi_{0}} \qquad$ & $ \overline{C} = \frac {C} {C_{moy}}  \qquad $ & $ \overline{\varphi} = \frac {\varphi} {\varphi_{0}}\qquad $ &$\overline{\rho Z} = \frac {\rho Z} {\phi_{4} F C_{moy}}  $ \\
$ \overline{p} = \frac {p} {F \varphi_{0} C_{moy}} \qquad $ & $ \overline{z} = \frac {z} {e} \qquad $&$\overline{\varepsilon}=\frac{\varepsilon \varphi_{0}}{e^{2} \phi_{4} FC_{moy}} 
\qquad $&$\overline{D}=\frac{D_{z}}{e \phi_{4} FC_{moy}}$ \\
\end{tabular}
\end{equation}

On obtient finalement le système suivant :
\begin{equation}
\begin{tabular}{ll}
$\overline{E}=-\frac{d\overline{\varphi }}{d\overline{z}} \qquad $& $\overline{D} = \overline{\varepsilon} \overline{E}$ \\ 
$\frac{d\overline{D}}{d\overline{z}}=\overline{\rho Z} \qquad $ &$\overline{\varepsilon} = A_{0} \overline{C} + A_{1}$\\
$\frac{d\overline{p}}{d\overline{z}}=\left( \overline{C}+A_{3}\right) \overline{E} \qquad \qquad$& $\frac{d\overline{C}}{d\overline{z}}=A_{2} \overline{C} \overline{E}$\\
$\overline{\rho Z}=\overline{C}-1$ & $\overline{D}(1)=\overline{D}(-1)$\\
$ \overline{\varphi }(-1)=1$&$\overline{\varphi }(1)=0$\\
\end{tabular}
\label{syst}
\end{equation}
avec :
\begin{equation*}
\begin{tabular}{llll}
$A_{0}=\frac{\varphi_{0} (\varepsilon_{moy}-\varepsilon^{0})}{e^{2}\phi_{4}FC_{moy}} \quad$ & $A_{1}=\frac{\varphi_{0} \varepsilon^{0}}{e^{2}\phi_{4}FC_{moy}} \quad$ & $A_{2}=\frac{F\varphi _{0}}{RT}\sim 38,7 \quad$ & $A_{3}=-\frac{\rho _{2}^{0}Z_{3}}{C_{moy}F}\sim 0,303$ \\ 
\end{tabular}%
\end{equation*}
On peut en tirer les relations suivantes :
\begin{equation}
\begin{tabular}{llll}
$\overline{C}=B_{1}\exp \left( -A_{2}\overline{\varphi }\right) \qquad \qquad $ &$\overline{p}=\frac{\overline{C}}{A_{2}}-A_{3}\overline{\varphi }+B_{2}$ \\
\end{tabular}%
\end{equation}
\begin{equation}
\frac{\overline{D}^{2}}{2} = \frac{1}{A_{2}} \left( \frac{A_{0}}{2} \overline{C}^{2}+(A_{1}-A_{0}) \overline{C} \right) + A_{1} \overline{\varphi} + \frac{1}{2A_{2}} \left( A_{0} - 2A_{1} - 2A_{1} \ln B_{1}\right)
\end{equation}
où, $B_{2}$ est une constante et où, si $\frac{2 A_{2} A_{1}}{A_{0}} <<1$, $B_{1}$ vérifie :
\begin{equation}
A_{0} (B_{1}-2) B_{1} \simeq 2 A_{1} (A_{2}-B_{1})
\end{equation}

La lame de polymère peut être assimilée à un matériau conducteur. On en déduit que le champ électrique est nul dans toute la lame excepté près des bords. Grâce au système d'équations précédent, on peut calculer les valeurs des différents paramètres au centre de la lame et aux extrémités ($\overline{p}$ est définie à une constante additive près) :
\begin{equation*}
\begin{tabular}{|c|c|c|c|}
\hline
& $-1$ & centre & $1$ \\ \hline
$\overline{C}$ & $B_{1}\exp \left( -A_{2}\right)$ & $1$ & $B_{1}$ \\ \hline
$\overline{\varphi }$ & $1$ & $\frac{\ln B_{1}}{A_{2}}$ & $0$ \\ \hline
$\overline{D}$ & $\sqrt{ \frac{1}{A_{2}} \left[  A_{0} + 2 A_{1} (A_{2}-1- \ln B_{1}) \right]} $ & $0$ & $\sqrt{ \frac{1}{A_{2}} \left[  A_{0} + 2 A_{1} (A_{2}-1- \ln B_{1}) \right]} $ \\ \hline
$\overline{E}$ & $\frac{1}{A_{1}} \sqrt{ \frac{1}{A_{2}} \left[  A_{0} + 2 A_{1} (A_{2}-1- \ln B_{1}) \right]} $ & $0$ & $\frac{1}{A_{0} B_{1} +A_{1}} \sqrt{ \frac{1}{A_{2}} \left[  A_{0} + 2 A_{1} (A_{2}-1- \ln B_{1}) \right]} $ \\ \hline
$\overline{\varepsilon}$ & $A_{1} $ & $A_{0}+A_{1}$ & $A_{0} B_{1} +A_{1} $ \\ \hline
$\overline{p}$ & $\frac{B_{1}\exp \left( -A_{2}\right) }{A_{2}}-A_{3}$ & $\frac{1}{A_{2}}-\frac{A_{3}\ln B_{1}}{A_{2}}$ & $\frac{B_{1}}{A_{2}}$ \\ \hline
$\overline{\rho Z}$ & $B_{1}\exp \left( -A_{2}\right)-1$ & $0$ & $B_{1}-1$ \\ \hline
\end{tabular}%
\end{equation*}
On peut alors faire une évaluation du terme de pression dans l'équation (\ref{Nernst}) et vérifier qu'il est bien négligeable, quel que soit le modèle de permittivité retenu (l'erreur commise est inférieure à $2 \%$ dans le cas nominal).

Le moment fléchissant est donné par :
\begin{equation}
\begin{tabular}{ll}
&$M^{p}_{A}=A_{5} \int_{-1}^{1} \overline{p} \overline{z} \;d\overline{z}
= A_{5} \int_{-1}^{1} (\frac{\overline{C}}{A_{2}}-A_{3}\overline{\varphi }) \overline{z} \;d\overline{z}$ \medskip\\
avec   &$A_{5}=6le^{2}F\varphi_{0}C_{moy} \simeq 0,045 \; Nm $\\
&$\int_{-1}^{1} \overline{C} \overline{z} \;d\overline{z}
= 2 \overline{D}(1) - \frac{A_{0}}{A_{2}} B_{1} - A_{1} $\\
\end{tabular}
\label{Mp}
\end{equation}

\section{Résolution avec différents modèles de permittivité}
\medskip
La résolution du système d'équation s'avère délicate d'un point de vue numérique en raison de la raideur des fonctions près des bords. 

Dans le cas d'une permittivité constante ($A_{0}=0$), $B_{1}=A_{2}$; on utilise pour résoudre la variable $y=\ln \overline{C}$ qui vérifie :
\begin{equation}
\begin{tabular}{l}
$y" = \frac{A_{2}}{A_{1}}(e^{y}-1)$\\
$y'(1)\simeq y'(-1)=\sqrt{2 \frac{A_{2}}{A_{1}}(A_{2}-ln A_{2}-1})  $\\
\end{tabular}
\label{y}
\end{equation}

Cette équation peut être résolue sous Matlab$^{\mbox{\scriptsize{\textregistered}}}$ en utilisant la subroutine \textit{bvp5c}. On en déduit $\overline{C}$, $\overline{E}$, $\overline{D}$, $\overline{\rho Z}$, $\overline{\varphi}$ et $\overline{p}$ à l'aide du système (\ref{syst}), puis $M_{A}^{p}$ par (\ref{Mp}) et enfin $w$, $F^{p}$ et $\theta$ grâce à (\ref{w}) et (\ref{FP}).

Dans le cas d'une permittivité linéaire ($A_{1}=0)$, $B_{1} \simeq 2$. La concentration en cations vérifie :
\begin{equation}
\frac{d^{2}\overline{C}}{d\overline{z}^{2}}=\frac{1}{\overline{z_{0}}^{2}} \left( \overline{C}-1\right)
\end{equation}
où $\overline{z_{0}}=\sqrt{\frac{A_{0}}{A_{2}}} \simeq 4,8 \; 10^{-3}$. Cette équation peut être résolue analytiquement :
\begin{equation}
\overline{C} = 1 + (B_{1}-1) \frac{\sinh\left( \overline{z}/\overline{z_{0}} \right)}{\sinh\left( 1/\overline{z_{0}} \right)}
\end{equation}
Les autres grandeurs sont calculées comme précédemment. Les fonctions obtenues étant particulièrement raides près des bords dans ce cas, il s'avère nécessaire de calculer l'intégrale de la pression par morceaux en utilisant des développements limités dans chaque intervalle.

Dans le cas affine, la constante $B_{1}$ vérifie :
\begin{equation}
B_{1}=\frac{A_{0}-A_{1}}{A_{0}} \left(1 + \sqrt{1 + \frac{2 A_{2} A_{0} A_{1}}{(A_{0} - A_{1})^{2}}} \right) \simeq 1 + \sqrt{1 + \frac{2 A_{2} A_{1}}{A_{0}}} \sim 2
\end{equation}
compte tenu des valeurs de permittivité choisies.
On utilise l'équation régissant $\overline{D}$ pour la résolution :
\begin{equation}
\frac{d^{2} \overline{D}}{d \overline{z}^{2}} = \frac{A_{2}\left( 1+ \frac{d \overline{D}}{d \overline{z}} \right)}{A_{0} \left( 1+ \frac{d \overline{D}}{d \overline{z}} \right) + A_{1}} \overline{D}
\end{equation}
équation résolue numériquement à l'aide de Matlab$^{\mbox{\scriptsize{\textregistered}}}$. Les quantités associées (pression, concentration, potentiel électrique...) sont obtenues comme dans les deux cas premiers cas.

\section{Résultats}
\medskip
\subsection{Profils des différentes grandeurs}
\medskip
Les profils des différentes grandeurs dans l'épaisseur ont été tracés dans le cas d'une lame de Nafion $Li^{+}$ correspondant aux données du paragraphe 4.

On constate que les profils de concentration, induction électrique, potentiel électrique et pression adimensionnés ont des aspects similaires : les courbes sont quasi constantes dans la zone centrale et varient très fortement à proximité des bords.

\begin{figure*} [h]
\includegraphics[width=0.9\textwidth]{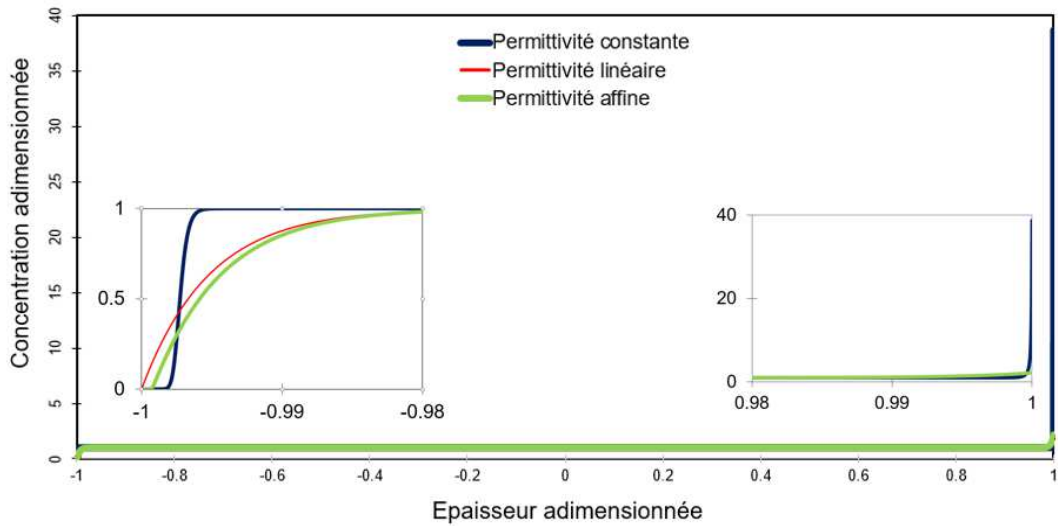}
\caption{Variation de la concentration en cations dans l'épaisseur de la lame}
\label{fig:4}
\end{figure*}

Les profils de concentration en cations fournis par les trois modèles de permittivité diffèrent près de la borne inférieure : on observe une zone presque totalement dépourvue de cations de $0,2 \; \mu m$ d'épaisseur dans le cas constant, et de $0,07 \; \mu m$ d'épaisseur dans le cas affine. Dans le cas linéaire, le profil de concentration a une pente très élevée sur l'électrode; il convient cependant de remarquer que le modèle n'est pas correct dans cette zone : en effet, la permittivité du matériau tend vers $0$ dans cette région, ce qui n'est pas acceptable physiquement.
Près de la borne supérieure, on observe une accumulation de cations sur une longueur caractéristique dépendant du modèle de permittivité choisi : $0,2 \; \mu m$ pour une permittivité constante, et $2 \; \mu m$ dans les cas linéaire et affine. La concentration sur l'électrode négative atteint par ailleurs une valeur 20 fois plus élevée dans le cas d'une permittivité constante que dans les deux autres cas.

\begin{figure*} [h]
\includegraphics[width=0.9\textwidth]{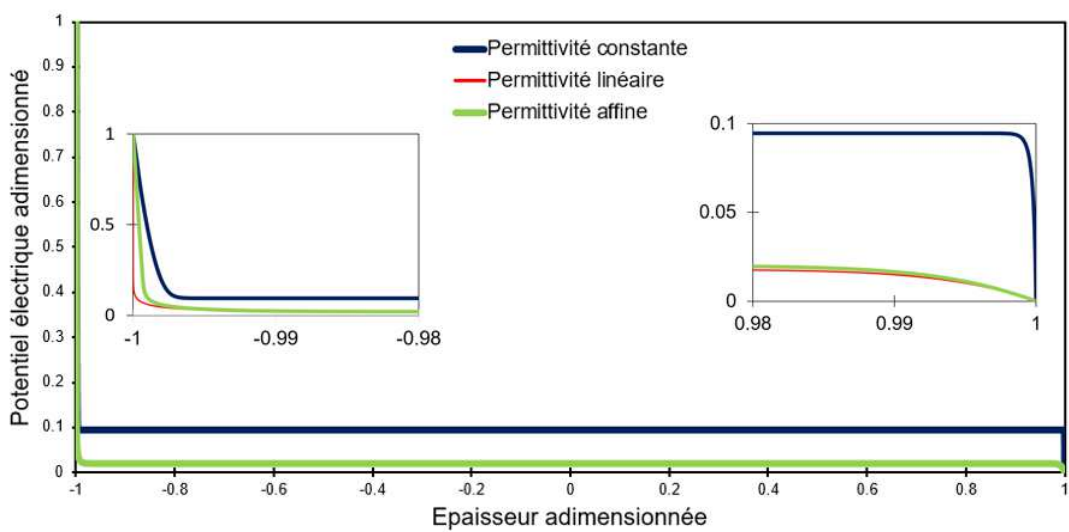}
\caption{Variation du potentiel électrique dans l'épaisseur de la lame}
\label{fig:5}
\end{figure*}

\begin{figure*} [h]
\includegraphics[width=0.9\textwidth]{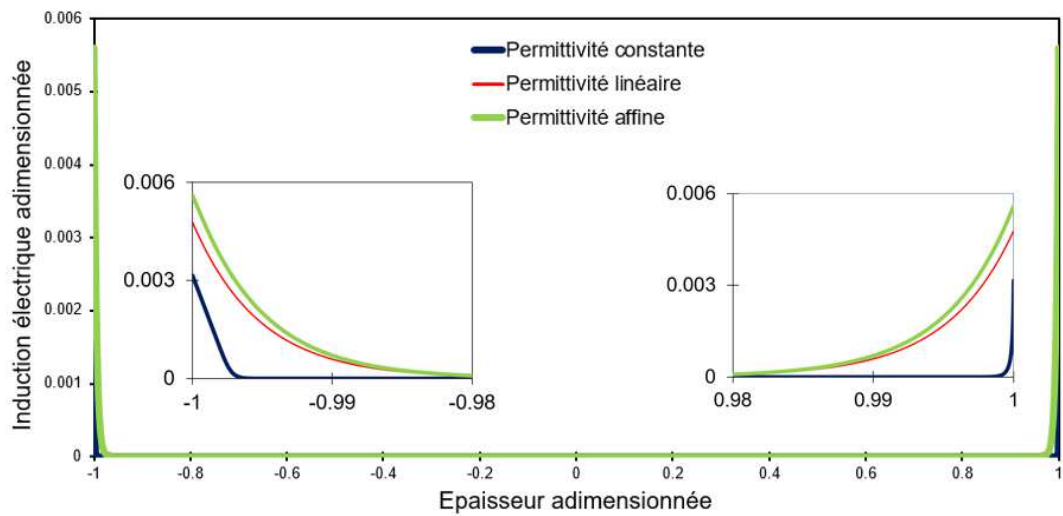}
\caption{Variation de l'induction électrique dans l'épaisseur de la lame}
\label{fig:6}
\end{figure*}

\begin{figure*} [h]
\includegraphics[width=0.9\textwidth]{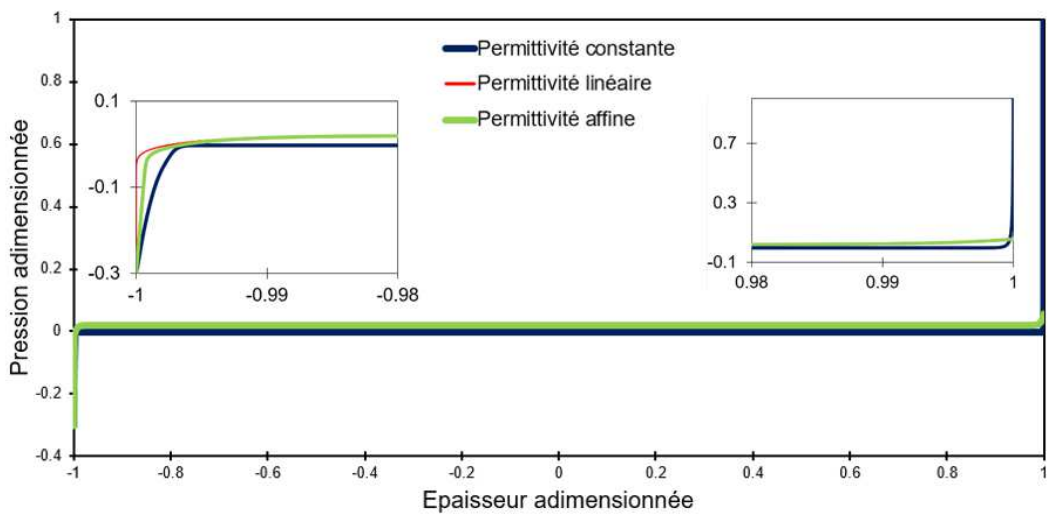}
\caption{Variation de la pression dans l'épaisseur de la lame}
\label{fig:7}
\end{figure*}

Les modèles linéaire et affine donnent des résultats très comparables pour les autres profils. Le modèle constant se distingue par sa raideur au voisinage des bornes.
Au voisinage de la borne $-1$, les longueurs caractéristiques sont de l'ordre de $0,2 \; \mu m$ pour les modèles constant et affine et pour les profils de pression et de potentiel électrique avec des valeurs aux bornes identiques. Les valeurs de l'induction électrique sont également très proches, mais la courbe obtenue avec le modèle affine est 5 fois moins raide qu'avec le modèle constant.
Au voisinage de la borne $+1$, le modèle constant fournit des profils d'induction, de potentiel électrique et de pression 20 fois plus raides environ que les deux autres modèles, avec des longueurs caractéristiques de l'ordre de quelques centièmes de micron. Les valeurs en $+1$ de l'induction électrique sont très voisines, en revanche la pression atteint une valeur beaucoup plus élevée avec le modèle constant. 

On vérifie en outre que l'induction électrique est quasiment nulle dans la partie centrale de la lame, ce qui corrobore le fait que le matériau se comporte comme un conducteur.

\subsection{Lois d'échelle}
\medskip
On vérifie numériquement qu'un calcul en grands déplacements est nécessaire dès que la flèche excède $3 \; mm$; pour la force de blocage, en revanche, l'écart entre les modèles en petits et en grands déplacements est inférieur à $1\%$ pour l'ensemble des simulations.

Les grandeurs mécaniques s'écrivent :%
\begin{equation}
\begin{tabular}{ll}
$\theta =-\frac{3}{2}\frac{FC_{moy}}{E}\varphi _{0}\;\frac{L}{e} \; I$ &$ $ \\ 
$w=\frac{2}{3}\frac{E}{FC_{moy}\varphi _{0}}\frac{e}{I}\left[ \cos \left( \frac{3}{2} \frac{FC_{moy}}{E}\frac{L}{e}\varphi _{0}I\right) -1\right] 
\simeq -\frac{3}{4}\frac{FC_{moy}}{E}\varphi _{0}\;\frac{L^{2}}{e} \; I$ & si $w$ est faible\\ 
$F^{p}=3FC_{moy}\varphi _{0}\;\frac{le^{2}}{L}I$ &$ $%
\end{tabular}%
\end{equation}
où $I = \int_{-1}^{1}\overline{p}\;\overline{z}\;d\overline{z}$ dépend du matériau choisi, du potentiel imposé $\varphi_{0}$ et de l'épaisseur $e$ de la lame, mais ni de sa longueur $L$ ni de sa largeur $l$. En faisant un développement limité de $I$ dans les différentes régions du profil (près des bornes et dans la région centrale), on peut montrer qu'en première approximation $I$ est inversement proportionnel à $e$. La dépendance avec le potentiel imposé est plus complexe et dépend du modèle de permittivité choisi; dans le cas d'une permittivité constante, $I$ varie approximativement comme $\varphi_{0}^{1/2}$.

On en déduit le tableau de variation suivant :
\begin{equation}
\begin{tabular}{|l|c|c|c|}
\hline
& $ L $&$l$&$e$\\
\hline
$\theta$& $ L $&$Cte$&$e^{-2}$\\
\hline
$F^{p}$& $ L^{-1} $&$l$&$e$\\
\hline
$w$& $ \simeq L^{2} $&$Cte$&$e^{-2}$\\
\hline
\end{tabular}
\end{equation}
On vérifie que la force de blocage varie linéairement avec la largeur, ce qui est en bon accord avec le résultat de Newbury et al \cite{Newbury2003}.

Compte tenu des formules précédentes, la force de blocage est proportionnelle à $1/L$, ce qui est en accord avec les résultats expérimentaux de Newbury et al (\cite{Newbury2003}).
De même, l'angle de rotation est proportionnel à la longueur de la poutre. La flèche en petits déplacement est proportionnelle à $L^{2}$. D'après nos simulations, cette dépendance est également vérifiée avec une bonne approximation par la flèche en grands déplacements : suivant le modèle de permittivité retenu, la loi de puissance approximant le mieux nos simulations a un exposant compris  entre $1,89$ et $1,95$, donc proche de $2$, avec un coefficient de corrélation supérieur à $0,999$ dans tous les cas. Ce résultat est en bon accord avec Shahinpoor (\cite{Shahinpoor1999}).

\begin{figure*} [h]
\includegraphics[width=0.95\textwidth]{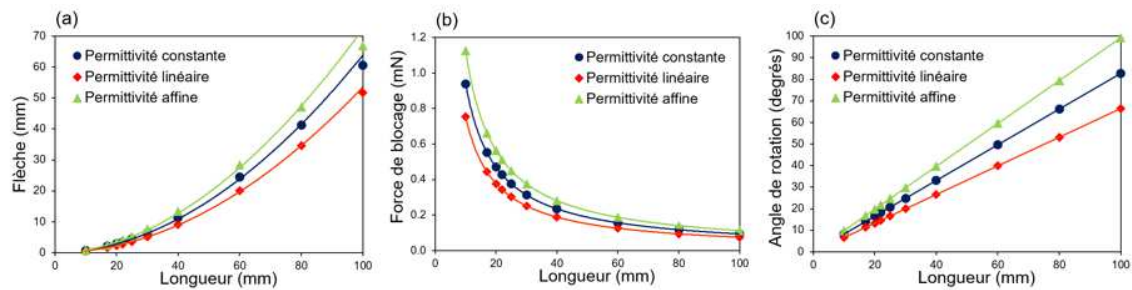}
\caption{Influence de la longueur : (a) Sur la flèche;  (b) Sur la force de blocage;  (c) Sur l'angle de rotation}.
\label{fig:8}
\end{figure*}

La force de blocage est proportionnelle à l'épaisseur. D'après nos simulations, la flèche varie quasiment comme $e^{-2}$ (on trouve un exposant compris entre $-1,9$ et $-1,95$ suivant le modèle de permittivité retenu, avec un coefficient de corrélation supérieur à $0,998$). Ce résultat est corroboré par les mesures de He et al \cite{He} ainsi que par les simulations de Vokoun et al \cite{Vokoun}. On observe également que la charge de l'électrode négative $FC(e)=FB_{1}C_{moy}$ est indépendante de l'épaisseur $e$ quel que soit le modèle de permittivité, ce qui est en accord avec les résultats obtenus par Lin \cite{Lin} pour un matériau voisin. 

\begin{figure*} [h]
\includegraphics[width=0.95\textwidth]{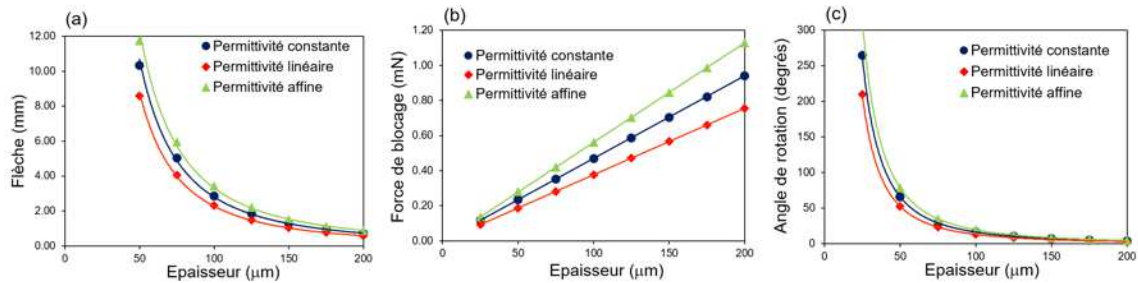}
\caption{Influence de l'épaisseur : (a) Sur la flèche;  (b) Sur la force de blocage;  (c) Sur l'angle de rotation}.
\label{fig:9}
\end{figure*}

\subsection{Influence du potentiel imposé}
\medskip
A la différence des lois d'échelle, la dépendance des différentes grandeurs avec $\varphi _{0}$ dépend du modèle de permittivité choisi.

Dans le cas d'une permittivité constante, on observe que la flèche varie à peu près linéairement avec le potentiel imposé (la corrélation est de $0,99$). On obtient un meilleur lissage avec une loi de puissance d'exposant $1,15$ (corrélation de $0,999$). La force de blocage et l'angle de rotation suivent approximativement la même tendance (corrélation de $0,955$ pour une loi linéaire et de $0,998$ pour une loi de puissance d'exposant $1,26$).
Le développement limité mentionné au paragraphe précédent indique une variation en $\varphi_{0}^{3/2}$ pour les trois grandeurs.
Une relation approximativement linéaire avait été observée expérimentalement dans le cas de l'effet inverse par Shahinpoor et al et Mojarrad (\cite{Shahinpoor1998}, \cite{Mojarrad}).

Dans le cas linéaire, en revanche, le moment est indépendant du potentiel imposé, de même que la flèche, la force de blocage et l'angle de rotation, ce qui est inacceptable physiquement.

Dans le cas affine, la corrélation avec une loi linéaire est mauvaise ($0,97$); il en va de même pour la force de blocage et l'angle de rotation (corrélation de $0,975$). On remarque en outre que les courbes ne tendent pas vers $0$ lorsque le potentiel imposé s'annule.

La variation des différentes grandeurs avec le potentiel électrique imposé s'avère donc discriminante pour le modèle de permittivité : seule une permittivité constante donne des résultats compatibles avec l'expérience.

\begin{figure*} [h]
\includegraphics[width=0.95\textwidth]{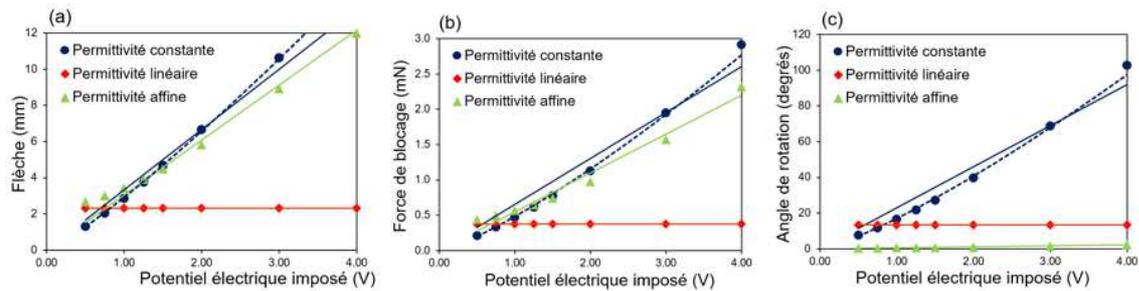}
\caption{Influence du potentiel électrique imposé : (a) Sur la flèche;  (b) Sur la force de blocage;  (c) Sur l'angle de rotation.}
\label{fig:10}
\end{figure*}

\section{Conclusion}
\medskip
Nous avons étudié la flexion d'une lame de polymère ionique à l'aide d'un modèle basé sur la thermomécanique des milieux continus que nous avions développé précédemment. La lame, encastrée à l'une de ses extrémités, est soumise à une différence de potentiel continue appliquée entre ses deux faces (cas statique). L'autre extrémité est soit libre, soit bloquée à l'aide d'un effort tranchant. Les grandeurs mécaniques (flèche, force de blocage et angle de rotation) ont été déterminés à l'aide d'un modèle de poutre en grands déplacements. Le matériau choisi pour effectuer les simulations est le Nafion. Trois modèles de permittivité (fonctions constante, linéaire et affine de la concentration en cations) sont étudiés.\newline
La résolution des équations de notre modèle nous a permis de tracer les profils de concentration en cations, pression, potentiel et induction électrique dans l'épaisseur de la lame. Ces grandeurs sont quasiment constantes dans la partie centrale, mais varient de façon drastique au voisinage des électrodes, ce qui est caractéristique d'un matériau conducteur. Les lois d'échelle obtenues pour la flèche et la force de blocage sont en bon accord avec les données expérimentales publiées dans la littérature : en particulier, la flèche varie comme le carré de la longueur de la lame et est inversement proportionnelle au carré de l'épaisseur; la force de blocage varie linéairement avec la largeur et l'épaisseur et est inversement proportionnelle à la longueur.
La variation des grandeurs mécaniques avec le potentiel imposé dépend du modèle de permittivité choisi; seul le modèle de permittivité constante fournit des résultats compatibles avec les données expérimentales et sera donc conservé pour la suite de nos travaux.\newline
Nous envisageons maintenant d'appliquer notre modèle à d'autres matériaux voisins et d'étudier d'autres configurations, notamment le cas d'une lame encastrée à ses deux extrémités. Nous nous intéresserons également à la modélisation de l'effet inverse (mode capteur). 

\section{Notations}
\medskip
Les indices $k=1,2,3,4$ désignent respectivement les cations, le solvant, le solide et la solution. Les quantités non indicées sont relatives au matériau complet.\newline
\newline
\noindent$C$ ($C_{moy}$) : concentration molaire en cations (relative à la phase
liquide);\newline
\noindent$D$ : coefficient de diffusion de masse des cations dans la phase
liquide;\newline
\noindent$\overrightarrow{D}$ : induction électrique;\newline
\noindent$e$ : demi épaisseur de la lame;\newline
\noindent$\overrightarrow{E}$ : champ électrique;\newline
\noindent$F$ : constante de Faraday;\newline
\noindent$\overrightarrow{F^{p}}$ : force de blocage;\newline
\noindent$G$, $\lambda $, $E$ : coefficients élastiques;\newline
\noindent$I^{p}$ : moment quadratique de la poutre;\newline
\noindent$K$ : perméabilité intrinsèque de la phase solide;\newline
\noindent$l$ : demi largeur de la lame;\newline
\noindent$L$ : longueur de la lame;\newline
\noindent$M_{k}$ : masse molaire du constituant $k$;\newline
\noindent$\overrightarrow{M^{p}}$ ($\overrightarrow{M_{A}^{p}}$) : moment fléchissant; \newline
\noindent$p$ : pression;\newline
\noindent$R$ : constante universelle des gaz parfaits;\newline
\noindent$T$ : température absolue;\newline
\noindent$v_{k}$ : volume molaire partiel du constituant $k$ (relatif à la phase liquide);\newline
\noindent$\overrightarrow{V}$ ($\overrightarrow{V_{k}}$): vitesses;\newline
\noindent$w$ ($w_{p}$): flèche de la poutre (flèche en petits déplacements);\newline
\noindent$Z$ ($Z_{k}$) : charge électrique massique;\newline
\noindent$\varepsilon $ : permittivité diélectrique;\newline
\noindent$\utilde{\epsilon }$ : tenseur des déformations;\newline
\noindent$\eta$ : viscosité dynamique de l'eau;\newline
\noindent$\theta$ : angle de rotation des sections droites de la poutre;\newline
\noindent$\lambda _{v}$\textit{, }$\mu _{v}$\textit{\ }: coefficients viscoelastiques;\newline
\noindent$\rho $ ($\rho _{k}^{0}$) : masse volumique;\newline
\noindent$\utilde{\sigma }$ : tenseur des contraintes 
totales (d'équilibre);\newline
\noindent$\varphi$ ($\varphi_{0}$): potentiel électrique;\newline
\noindent$\phi _{k}$ : fraction volumique de la phase $k$;\newline


\begin{thebibliography}{9}
\bigskip
\bibitem{nemat2000} S. Nemat-Nasser, J. Li, Electromechanical response of
ionic polymers metal composites, Journal of Applied Physics, 87 (2000) 3321--3331.
%
\bibitem{Ishii06} M. Ishii, T. Hibiki, Thermo-fluid dynamics of two-phase
flow, Springer, New-York, 2006.
%
\bibitem{Tixier1} M. Tixier, J. Pouget, Conservation laws of an
electro-active polymer, Continuum Mechanics and Thermodynamics 26, 4 (2014) 465--481.
%
\bibitem{Tixier2} M. Tixier, J. Pouget, Constitutive equations for an electroactive polymer, Continuum 
Mechanics and Thermodynamics, 28, 4 (2016) 1071--1091.
%
\bibitem{Collette} F. Collette, Vieillissement hygrothermique du Nafion, Thèse, Université de Grenoble  I, 2008.

\bibitem{Satterfield2009} Barclay Satterfield M., Benziger J. B., Journal of polymer Science, Part B, 47, 11-24 (2009)
%
\bibitem{Nemat2002} S. Nemat-Nasser, Micromechanics of actuation of ionic polymer-metal composites, Journal 
of Applied Physics, 92, 5 (2002) 2899--2915.
%
\bibitem{NewburyTh} K.M. Newbury, Characterization, modeling and control of ionic-polymer transducers, 
Thesis, Faculty of the Virginia Polytechnic Institute and State University, Blacksburg, Virginia, 2002.
%
\bibitem{Newbury2002} K.M. Newbury, D.J. Leo, Linear Electromechanical modeling and characterization of ionic 
polymer benders, Journal of Intelligent Material Systems and Structures, 13 (2002) 51--60.

\bibitem{Newbury2003} K.M. Newbury, D.J. Leo, Electromechanical Model of ionic polymer transducers - Part II : 
experimental validation, Journal of Intelligent Material Systems and Structures, 14 (2003) 343--357.

\bibitem{Deng} Z.D. Deng, K.A. Mauritz, Dielectric relaxation studies of water-containing short side chain 
perfluorosulfonic acid membranes, Macromolecules, 25 (1992) 2739--2745.

\bibitem{Farinholt} Farinholt K., Leo D.J., Mechanics of Materials, 36,421-433 (2004)
%
\bibitem{Shahinpoor1999} M. Shahinpoor, Electro-mechanics of iono-elastic beams as electrically-controllable artificial muscles, Proceedings of SPIE, 3669 (1999) 109--121.

\bibitem{He} Qingsong He, Min Yu, Linlin Song, Haitao Ding, Xiaoqing Zhang, Zhendong Dai, "Experimental Study and Model Analysis of the Performance of IPMC Membranes with Various Thickness", Journal of Bionic Engineering, Volume 8, 2011, Pages 77-85

\bibitem{Vokoun} David Vokoun, Qingsong He, Ludek Heller, Min Yu, Zhendong Dai, "Modeling of IPMC Cantilever's Displacements and Blocking Forces", Journal of Bionic Engineering, Volume 12, Issue 1, January 2015, Pages 142-151

\bibitem{Lin} Lin J.H., Liu Y., Zhang Q.M., "Influence of the Electrolyte Film Thickness on Charge Dynamics of Ionic Liquids in Ionic Electroactive Devices", Macromolecules, 2012, 45 (4), pp 2050--2056
%
\bibitem{Shahinpoor1998} M. Shahinpoor, Y. Bar-Cohen, J.O. Simpson, J. Smith, Ionic polymer-metal composites
(IPMCs) as biomimetic sensors, actuators and artificial muscles; a review, Smart Materials and Structures 7, 
(1998) 15--30.

\bibitem{Mojarrad} M. Mojarrad, M. Shahinpoor, Ion-exchange-metal composite sensor films, Proceedings of the 
SPIE, 3042 (1997) 52--60.

\end{thebibliography}
\end{document}